\documentclass[12pt]{article}

\usepackage{latexsym,wrapfig,epsfig,url}

\newcommand{\new}{}

\def\Clobber{{\sc Solitaire-Clobber}}
\def\Hamgrid{{\sc Grid-Hamiltonicity}}
\def\Right{matching}
\def\Wrong{clashing}
\def\pot{\delta}

\def\etal{{\it et al.}}
 
\long\def\ignore#1{}
\let\ts\thinspace

\def\inst#1{$^{#1}$}
\def\institute#1{
  \maketitle
  \global\let\maketitle\relax
  \def\and{\\ $^2$\quad
     \gdef\and{\\ $^3$\quad}
  }
  \begin{minipage}[c]{.8\textwidth}\begin{center} $^1$\quad #1 \end{center}\end{minipage}
  \bigskip
}

\advance \topmargin by -\headheight
\advance \topmargin by -\headsep
\evensidemargin \oddsidemargin
\let\latexcite=\cite
\def\cite{\nolinebreak\latexcite}
\let\latexref=\ref
\def\ref{\nolinebreak\latexref}

\newtheorem{theorem}{Theorem}

\newtheorem{corollary}[theorem]{Corollary}

\newenvironment{proof}{\textbf{Proof: }%
  \gdef\ProofEnding{\ifmmode ~~\Box
                    \else \hspace*{\fill}$\Box$\par \medskip
                    \fi \ignorespaces}}
  {\ProofEnding}

{\makeatletter
 \gdef\xxxmark{%
   \expandafter\ifx\csname @mpargs\endcsname\relax 
     \expandafter\ifx\csname @captype\endcsname\relax 
       \marginpar{xxx}
     \else
       xxx 
     \fi
   \else
     xxx 
   \fi}
 \gdef\xxx{\@ifnextchar[\xxx@lab\xxx@nolab}
 \long\gdef\xxx@lab[#1]#2{{\bf [\xxxmark #2 ---{\sc #1}]}}
 \long\gdef\xxx@nolab#1{{\bf [\xxxmark #1]}}
}

\def\alternating#1{A_{#1}}
\def\whether#1{\{1\hbox{ if }#1\}}
\def\X{\ensuremath{\bullet}}
\def\O{\ensuremath{\circ}}
\newlength\blanklength
\newlength\blankheight
\def\B{\parbox[c][\blankheight]{\blanklength}{\centering .}}

\def\pmod#1{\ (\mathrm{mod}\ #1)}
\long\def\conf#1{\ensuremath{\vcenter{%
  \def\\{\unskip\egroup\vskip -6pt\hbox\bgroup\ignorespaces}%
  \hbox\bgroup\ignorespaces #1\unskip\egroup}}}
\def\move{\ts\ensuremath{\to}\ts}
\def\jumpw#1{\ts\ensuremath{\stackrel{#1}{\to}}\ts}
\def\dstackrel#1#2{\mathrel{\mathop{#2}\limits_{#1}}}
\def\jumpb#1{\ts\ensuremath{\dstackrel{#1}{\to}}\ts}
\long\def\seq#1{\begin{center}#1\end{center}}

\def\lparen{\ts\ensuremath{\bigg(}\ts}
\def\rparen{\ts\ensuremath{\bigg)}\ts}

\begin{document}

\title{Solitaire Clobber}
\author{Erik D. Demaine\inst{1} \and Martin L. Demaine\inst{1}
  \and Rudolf Fleischer\inst{2}}
\institute{%
     MIT Laboratory for Computer Science, Cambridge, MA 02139, USA,
     email: \{\texttt{edemaine}, \texttt{mdemaine}\}\texttt{@mit.edu}
  \and HKUST Department of Computer Science, Clear Water Bay, Kowloon,
     Hong Kong, email: \texttt{rudolf@cs.ust.hk}}
\date{}


\begin{abstract}
Clobber is a new two-player board game.
In this paper, we introduce the one-player variant Solitaire Clobber
where the goal is to remove as many stones as possible from the board
by alternating white and black moves.
We show that a checkerboard configuration on a single row
(or single column)
can be reduced to about $n/4$ stones.
For boards with at least two rows and \new two columns,
we show that a checkerboard configuration can be reduced to
a single stone if and only if the number of stones is not a multiple of
three, and otherwise it can be reduced to two stones.
\new We also show that in general
it is NP-complete to decide whether an arbitrary
Clobber configuration can be reduced to a single stone.
\end{abstract}

\section{Introduction}
\label{s_intro}

Clobber is a new two-player combinatorial board game with complete
information, recently introduced by Albert, Grossman, and Nowakowski
(see \cite{clobber_www}).
It is played with black and white stones
occupying some subset of the squares of an $n\times m$ checkerboard.
The two players, White and Black, move alternately by
picking up one of their own stones and {\em clobbering}
an opponent's stone on a horizontally or vertically adjacent square.
The clobbered stone is removed from the board and replaced by the stone
that was moved.
The game ends when one player, on their turn, is unable to move,
and then that player loses.

We say a stone is {\em \Right} if it has the same color as the square it
occupies on the underlying checkerboard; otherwise it is
{\em \Wrong}.
In a {\em checkerboard configuration}, all stones are \Right,
i.e., the white stones occupy white squares and the black
stones occupy black squares.
And in a {\em rectangular} configuration, the stones occupy exactly
the squares of some rectangular region on the board.
Usually, Clobber starts from a rectangular
checkerboard configuration, and White moves first
(if the total number of stones is odd we assume that it is
White who has one stone less than Black).

At the recent Dagstuhl Seminar on Algorithmic Combinatorial
Game Theory \cite{dag02}, the game was first introduced to
a broader audience.
Tom\'a\v{s} Tich\'y from Prague won the first Clobber tournament,
played on a $5\times 6$ board, beating his supervisor
Ji\v{r}\'{\i} Sgall in the finals.
Not much is known about Clobber strategies,
even for small boards,
and the computation of CGT game values is also only
in its preliminary stages.

\medskip

In this paper we introduce {\em Solitaire Clobber},
where a single player (or two cooperative players)
tries to remove as many
stones as possible from the board by alternating
white and black moves.
If the configuration ends up with $k$ immovable stones,
we say that the initial board configuration is {\em reduced} to $k$ stones,
or {\em $k$-reduced}.
Obviously, 1-reducibility can only be possible
if half of the stones are white (rounded down),
and half of the stones are black (rounded up).
But even then it might not be possible.

We prove the following necessary condition
for a Clobber position to be 1-reducible:
The number of stones plus the number of \Wrong\ stones
cannot be a multiple of three.
Surprisingly, this condition is also sufficient
for truly two-dimensional rectangular checkerboard configurations
(i.e., with at least two rows and two columns).
And if the condition is not true, then the board is 2-reducible
(with the last two stones separated by a single empty square),
which is the next-best possible.
However, in general, we show that it is NP-complete to decide whether an
arbitrary \new(non-rectangular non-checkerboard)
configuration is 1-reducible.

\new In one-dimensional Solitaire Clobber (i.e., the board consists of
a single row of stones) we can usually not expect 1-reducibility.
We show that the checkerboard configuration can be reduced
to $\lceil n/4 \rceil + \whether{n \equiv 3 \pmod{4}}$ stones,
no matter who moves first, and that this bound is best possible
even if we do not have to alternate between white and black moves.
This result was obtained independently by Grossman \cite{Gro02}.

\medskip

This paper is organized as follows.
In Section \ref{s_1dclobber}, we analyze the reducibility
of \new one-dimensional checkerboard configurations.
In Section~\ref{s_2dclobber}, we study reducibility of
\new truly two-dimensional rectangular checkerboard configurations. 
And in Section~\ref{s_npc} we show that deciding 1-reducibility
is NP-complete \new for general configurations.
We conclude with some open problems in Section~\ref{s_conclusions}.

\section{One-Dimensional Solitaire Clobber}
\label{s_1dclobber}

In this section we study Solitaire Clobber played on a board
consisting of a single row of stones.
Let $\alternating{n}$ denote the checkerboard configuration,
i.e., an alternating sequence of white and black stones.
By symmetry, we can assume throughout this section
that $\alternating{n}$ always starts with
a black stone, so we have $\alternating{n}=\hbox{\Large \X\O\X\O}\cdots$.
We first show an upper bound on the $k$-reducibility of checkerboard
configurations.

\begin{theorem}
\label{thm_1dclob_up}
For $n\ge 1$,
the configuration $\alternating{n}$
can be reduced to
$\lceil n/4 \rceil + \whether{n \equiv 3 \pmod{4}}$ stones
by an alternating sequence of moves,
no matter who is to move first.
\end{theorem}

\begin{proof}
Split the configuration $\alternating{n}$
into $\lceil n/4 \rceil$ substrings, all but possibly one of length
four.
Each substring of length one, two, or four
can be reduced to one stone by
alternating moves, no matter which color moves first.
And a substring of size three can be reduced to two stones by one move,
no matter which color moves first.
\end{proof}

In this move sequence, we end up with one isolated
stone somewhere in the middle of each block of four consecutive stones.
One might wonder whether a more clever strategy could end
up with one stone at the end of each subblock,
and then we could clobber one more stone in each pair of adjacent
stones from the subblocks.
Unfortunately, this is not possible, as shown by the following
matching lower bound.
The lower bound holds even if we are not forced to alternate between
white and black moves.
We give a simple proof for the theorem due to Grossman~\cite{Gro02}.

\begin{theorem}
\label{thm_1dclob_low}
Let $n\ge 1$.
Even if we are not restricted to alternating white and black moves,
the configuration $\alternating{n}$ cannot be reduced to fewer than
$\lceil n/4 \rceil + \whether{n \equiv 3 \pmod{4}}$ stones.
\end{theorem}

\begin{proof}
First, it is not possible to reduce \new $\alternating{k}$
to a single stone, for $k=3$ or $k\ge 5$.
Second, each stone in the final configuration comes from some contiguous
substring of stones in the initial configuration.
But each of these substrings can have only one, two, or four stones.
Thus, there are at least $\lceil n/4 \rceil$ stones left at the end,
and even one more if $n \equiv 3 \pmod{4}$.
\end{proof}

Somewhat surprisingly, the tight bound of Theorems \ref{thm_1dclob_up}
and \ref{thm_1dclob_low} is not monotone in $n$, the number of
stones in the initial configuration.  See Table~\ref{t_1dclobber}.

\begin{table}[h]
\centering
\begin{tabular}{llc}
\multicolumn{2}{l}{Configuration} & Reducibility \\ \hline
$\alternating{1}$  & \Large \X & 1 \\
$\alternating{2}$  & \Large \X\O & 1 \\
$\alternating{3}$  & \Large \X\O\X & 2 \\
$\alternating{4}$  & \Large \X\O\X\O & 1 \\
$\alternating{5}$  & \Large \X\O\X\O\X & 2 \\
$\alternating{6}$  & \Large \X\O\X\O\X\O & 2 \\
$\alternating{7}$  & \Large \X\O\X\O\X\O\X & 3 \\
$\alternating{8}$  & \Large \X\O\X\O\X\O\X\O & 2 \\
$\alternating{9}$  & \Large \X\O\X\O\X\O\X\O\X & 3 \\
$\alternating{10}$ & \Large \X\O\X\O\X\O\X\O\X\O & 3 \\
$\alternating{11}$ & \Large \X\O\X\O\X\O\X\O\X\O\X & 4 \\
$\alternating{12}$ & \Large \X\O\X\O\X\O\X\O\X\O\X\O & 3 \\
\end{tabular}
\caption{\label{t_1dclobber}
  Reducibility of one-dimensional checkerboard Clobber configurations.}
\end{table}

\section{Rectangular Solitaire Clobber}
\label{s_2dclobber}

In this section we study reducibility of rectangular checkerboard
configurations with at least two rows and two columns.
We first show a general lower bound on the reducibility that
holds for arbitrary Clobber configurations.
For a configuration $C$, we denote the quantity
``number of stones plus number of \Wrong\ stones'' by $\pot(C)$.

As it turns out, $\pot(C)\pmod{3}$ actually divides
all clobber configurations into three equivalence classes.
Any configuration will stay in the same equivalence class,
after any number of moves.
Because one of the three equivalence classes
(with $\delta(C)\equiv 0 \pmod{3}$) does not contain
configurations with a single stone, all configurations
in this equivalence class are not 1-reducible.
As in the 1-dimensional lower bound of
Theorem~\ref{thm_1dclob_low}, this is true even if
we allow arbitrary non-alternating move sequences.

\begin{theorem}
\label{thm_delta}
For a configuration $C$, $\delta(C)\pmod{3}$ does not change
after an arbitrary move sequence.
\end{theorem}

\begin{proof}
If we move a \Right\ stone in $C$ then $\pot$ drops by one
because we clobber another \Right\ stone, and $\pot$ rises by one
because our stone becomes \Wrong, so $\pot$ actually does not
change in this move.
If we move a \Wrong\ stone then $\pot $ drops by two
because we clobber another \Wrong\ stone, and $\pot$ drops by another one
because our stone becomes \Right, resulting in a total drop of three
for the move.
\end{proof}

\begin{corollary}
\label{thm_2dclob_low}
A configuration $C$ with $\delta(C)\equiv 0 \pmod{3}$ is not
1-reducible.
\end{corollary}

\begin{proof}
A single stone can only have $\pot$ equal to one or two
(depending on whether it is a \Right\ or \Wrong\ stone).
Thus, by the previous theorem,
configurations $C$ with $\pot(C)\equiv 0 \pmod{3}$ are not
1-reducible.
\end{proof}

The rest of this section is devoted to a proof that this bound
is actually tight for rectangular checkerboard configurations:

\begin{theorem}
\label{thm_2dclob_up}
For $n,m\ge 2$, a rectangular checkerboard configuration
with $n$ rows and $m$ columns is 2-reducible if
$nm \equiv 0 \pmod{3}$, and 1-reducible otherwise.
\end{theorem}

We present an algorithm that computes a sequence of moves that reduces
the given checkerboard configuration to one or two
stones as appropriate.

We distinguish cases in a somewhat complicated way.
There are finitely many cases with $2 \leq n,m \leq 6$;
these cases can be verified trivially, as shown in Appendix~\ref{appendix}.
The remaining cases have at least one dimension with at least seven
stones;
by symmetry, we ensure that the configuration has at least
seven columns.
These cases we distinguish based on the parities of $n$ and $m$:

  \begin{itemize}
  \item \textbf{Case EE:} Even number of rows and columns
    [Section \ref{ss_ee}]
  \item \textbf{Case OE:} Odd number of rows, even number of columns
    [Section \ref{ss_oe}]
  \item \textbf{Case EO:} Even number of rows, odd number of columns
    [Section \ref{ss_o}]
  \item \textbf{Case OO:} Odd number of rows and columns
    [Section \ref{ss_o}]
  \end{itemize}

Cases OE and EO are symmetric for configurations with at least
seven rows and at least seven columns.
By convention, we handle such situations in Case EO.
But when one dimension is smaller than seven, we require that
dimension to be rows, forcing us into Case OE or Case EO
and breaking the symmetry.
In fact, we solve these instances of Case OE by rotating the board and
solving the simpler cases E3 and E5 (even number of rows, and three
or five columns,
respectively).

Section \ref{ss_general} gives an overview of our general approach.
Section \ref{ss_ee} considers Case EE,
which serves as a representative example of the procedure.
Section \ref{ss_oe} extends this reduction to Case OE
(when the number of rows is less than seven),
which is also straightforward.
Finally, Section \ref{ss_o} considers the remaining more-tedious cases
in which the number of columns is odd.

\subsection{General Approach}
\label{ss_general}

In each case, we follow the same basic strategy.  We eliminate the stones on
the board from top to bottom, two rows at a time.  More precisely, each
\emph{step} reduces the topmost two rows down to $O(1)$
stones (usually one or two) arranged in a fixed pattern that
touches the rest of the configuration through the bottom row.

There are usually four types of steps, repeated in the order
$$ (1),\,\underbrace{(2),\,(3),\,(4)},\,
         \underbrace{(2),\,(3),\,(4)},\,
         \underbrace{(2),\,(3),\,(4)},\,\dots. $$
Step~(1) leaves a small remainder of stones from the top two rows
in a fixed pattern.
Step~(2) absorbs this remainder and the next two rows, in total reducing the
top four rows down to a different pattern of remainder stones.
Step~(3) leaves yet another pattern of remainder stones from the top six rows.
Finally, step~(4) leaves the same pattern of remainder stones from step (1),
so the sequence can repeat (2), (3), (4), (2), (3), (4), \dots.

In some simple cases, steps~(1) and~(2) leave the same pattern of remainder
stones.  Then just two types of steps suffice, repeating in the order
$(1), (2), (2), (2), \dots$.
In other cases, three steps suffice.

In any case, the step sequence may terminate with any type of step.
Thus, we must also show how to reduce each pattern of remainder stones
down to one or two stones as appropriate;
when needed, these final reductions are enclosed by parentheses
because they are only used at the very end.
In addition, if the total number of rows is odd, the final step involves three
rows instead of two rows, and must be treated specially.

In the description below, a single move is denoted by \move. 
But we often do not show long move sequences completely.
Instead, we usually `jump' several moves at a time,
denoted by \jumpw{a}\ or \jumpb{a},
depending on whether White or Black moves first,
where $a$ denotes the number of moves we jump.

\subsection{Case EE: Even Number of Rows and Columns}
\label{ss_ee}

We begin with the case in which both $n$ and $m$ are even.  This case is easier
than the other cases: the details are fairly clean.  It serves as a
representative example of the general approach.

Because the number of columns is even and at least seven,
it must be at least eight.
Every step begins by reducing the two involved rows
down to a small number of columns.
First, we clobber a few stones to create the following configuration
in which the lower row has two more stones than the upper row,
one on each side:
\seq{
\conf{\X\O\\ \O\X}$\cdots$\conf{\X\O\\ \O\X}
\jumpw{3}
\conf{\X\O\\ \O\X}$\cdots$\conf{\B\B\\ \O\B}
\jumpb{3}
\conf{\B\B\\ \B\X}$\cdots$\conf{\B\B\\ \O\B}
}
Then we repeatedly apply the following reduction,
in each step removing six columns, three on each side:
\seq{
\conf{\B\X\O\X\\ \X\O\X\O}$\cdots$\conf{\O\X\O\B\\ \X\O\X\O}
\jumpw{4}
\conf{\B\O\B\X\\ \B\X\X\O}$\cdots$\conf{\O\B\X\B\\ \X\O\O\B}
\jumpw{4}
\conf{\B\B\B\X\\ \B\B\O\O}$\cdots$\conf{\O\B\B\B\\ \X\X\B\B}
\jumpw{2}
\conf{\B\B\B\B\\ \B\B\B\O}$\cdots$\conf{\B\B\B\B\\ \X\B\B\B}
}
We stop applying this reduction when the bottom row has just six,
eight, or ten columns left, and the top row has four, six, or eight
columns, depending
on whether $m \equiv 2$, $1$, or $0 \pmod{3}$, respectively.

The resulting two-row configuration has either a black stone in the lower-left
and a white stone in the lower-right \new corner
$\Big(\conf{\B\X\\ \X\O}\cdots\conf{\O\B\\ \X\O}\Big)$, or vice versa
$\Big(\conf{\B\O\\ \O\X}\cdots\conf{\X\B\\ \O\X}\Big)$.
We show reductions for the former case; the latter case is symmetric.

\paragraph{Case 1: $m\equiv 2\pmod{3}$}

\begin{itemize}
\item[(1)]

\quad
\conf{\B\X\O\X\O\B\\ \X\O\X\O\X\O}
\jumpw{3}
\conf{\B\B\B\X\O\B\\ \B\O\X\O\X\O}
\jumpb{3}
\conf{\B\B\B\B\B\B\\ \B\O\X\O\X\B}
\jumpw{2}
\conf{\B\B\B\B\B\B\\ \B\B\O\X\B\B}
\lparen
\move
\conf{\B\B\B\B\B\B\\ \B\B\B\O\B\B}
\rparen
\raisebox{6pt}{\footnotemark[1]}
\footnotetext[1]{Parenthetical moves are made only if this is the final step.}

\item[(2)]
\quad
\conf{\B\B\O\X\B\B\\ \B\X\O\X\O\B\\ \X\O\X\O\X\O}
\jumpw{2}
\conf{\B\B\O\B\B\B\\ \B\X\O\X\B\B\\ \X\O\X\O\X\O}
\jumpw{2}
\conf{\B\B\O\B\B\B\\ \B\B\X\X\B\B\\ \X\O\X\O\O\B}
\jumpw{2}
\conf{\B\B\B\B\B\B\\ \B\B\O\X\B\B\\ \B\X\X\O\O\B}
\jumpw{2}
\conf{\B\B\B\B\B\B\\ \B\B\B\B\B\B\\ \B\X\O\X\O\B}
\jumpw{3}
\conf{\B\B\B\B\B\B\\ \B\B\B\B\B\B\\ \B\B\O\B\B\B}

\item[(3)]
\quad
\conf{\B\B\O\B\B\B\\ \B\X\O\X\O\B\\ \X\O\X\O\X\O}
\jumpb{2}
\conf{\B\B\O\B\B\B\\ \B\X\O\X\O\B\\ \B\X\X\O\O\B}
\jumpb{2}
\conf{\B\B\B\B\B\B\\ \B\B\O\X\O\B\\ \B\X\X\O\O\B}
\jumpb{2}
\conf{\B\B\B\B\B\B\\ \B\B\B\B\X\B\\ \B\X\O\O\O\B}
\jumpb{1}
\conf{\B\B\B\B\B\B\\ \B\B\B\B\B\B\\ \B\X\O\O\X\B}
\jumpw{2}
\conf{\B\B\B\B\B\B\\ \B\B\B\B\B\B\\ \B\O\B\X\B\B}

\item[(4)]
\quad
\conf{\B\O\B\X\B\B\\ \B\X\O\X\O\B\\ \X\O\X\O\X\O}
\jumpw{2}
\conf{\B\O\B\B\B\B\\ \B\X\B\X\O\B\\ \X\O\X\O\X\O}
\jumpw{2}
\conf{\B\B\B\B\B\B\\ \B\O\B\B\X\B\\ \X\O\X\O\X\O}
\jumpw{2}
\conf{\B\B\B\B\B\B\\ \B\O\B\B\X\B\\ \B\X\X\O\O\B}
\jumpw{2}
\conf{\B\B\B\B\B\B\\ \B\B\B\B\B\B\\ \B\O\X\O\X\B}
\jumpw{2}
\conf{\B\B\B\B\B\B\\ \B\B\B\B\B\B\\ \B\B\O\X\B\B}
\lparen
\move
\conf{\B\B\B\B\B\B\\ \B\B\B\B\B\B\\ \B\B\B\O\B\B}
\rparen
\end{itemize}

\paragraph{Case 2: $m\equiv 1\pmod{3}$}

\begin{itemize}
\item[(1)]

First we clear another six columns and obtain

\quad
\conf{\B\X\O\X\O\X\O\B\\ \X\O\X\O\X\O\X\O}
\jumpw{12}
\conf{\B\B\B\B\B\B\B\B\\ \B\B\B\O\X\B\B\B}
\move
\conf{\B\B\B\B\B\B\B\B\\ \B\B\B\B\O\B\B\B}

\item[(2)]

\quad
\conf{\B\B\B\B\O\B\B\B\\ \B\X\O\X\O\X\O\B\\ \X\O\X\O\X\O\X\O}
\jumpb{3}
\conf{\B\B\B\B\O\B\B\B\\ \B\X\O\X\O\B\B\B\\ \X\O\X\O\X\O\X\B}
\jumpw{3}
\conf{\B\B\B\B\O\B\B\B\\ \B\B\B\X\O\B\B\B\\ \B\O\X\O\X\O\X\B}
\jumpb{2}
\conf{\B\B\B\B\B\B\B\B\\ \B\B\B\B\O\B\B\B\\ \B\O\X\O\X\O\X\B}
\jumpb{2}
\conf{\B\B\B\B\B\B\B\B\\ \B\B\B\B\B\B\B\B\\ \B\O\X\O\O\X\B\B}
\jumpb{4}
\conf{\B\B\B\B\B\B\B\B\\ \B\B\B\B\B\B\B\B\\ \B\B\B\O\B\B\B\B}

\item[(3)]

\quad
\conf{\B\B\B\O\B\B\B\B\\ \B\X\O\X\O\X\O\B\\ \X\O\X\O\X\O\X\O}
\jumpb{3}
\conf{\B\B\B\O\B\B\B\B\\ \B\X\O\X\O\B\B\B\\ \X\O\X\O\X\O\X\B}
\jumpw{2}
\conf{\B\B\B\O\B\B\B\B\\ \B\O\B\X\O\B\B\B\\ \B\X\X\O\X\O\X\B}
\jumpw{2}
\conf{\B\B\B\O\B\B\B\B\\ \B\B\B\X\O\B\B\B\\ \B\O\X\O\X\X\B\B}
\jumpw{2}
\conf{\B\B\B\O\B\B\B\B\\ \B\B\B\X\B\B\B\B\\ \B\O\X\O\X\B\B\B}
\jumpw{4}
\conf{\B\B\B\B\B\B\B\B\\ \B\B\B\B\B\B\B\B\\ \B\O\B\X\B\B\B\B}

\item[(4)]

\quad
\conf{\B\O\B\X\B\B\B\B\\ \B\X\O\X\O\X\O\B\\ \X\O\X\O\X\O\X\O}
\jumpw{2}
\conf{\B\O\B\X\B\B\B\B\\ \B\X\O\X\O\B\X\B\\ \X\O\X\O\X\O\O\B}
\jumpw{4}
\conf{\B\B\B\X\B\B\B\B\\ \B\B\O\X\O\B\B\B\\ \B\O\X\O\X\O\X\B}
\jumpw{2}
\conf{\B\B\B\B\B\B\B\B\\ \B\B\B\X\O\B\B\B\\ \B\O\X\O\X\O\X\B}
\jumpw{4}
\conf{\B\B\B\B\B\B\B\B\\ \B\B\B\B\O\B\B\B\\ \B\B\B\O\X\X\B\B}
\jumpw{3}
\conf{\B\B\B\B\B\B\B\B\\ \B\B\B\B\B\B\B\B\\ \B\B\B\B\O\B\B\B}
\end{itemize}

\paragraph{Case 3: $m\equiv 0\pmod{3}$}

\begin{itemize}
\item[(1)]

First we clear another six columns and obtain

\quad
\conf{\B\X\O\X\O\X\O\X\O\B\\ \X\O\X\O\X\O\X\O\X\O}
\jumpw{12}
\conf{\B\B\B\B\O\X\B\B\B\B\\ \B\B\B\O\X\O\X\B\B\B}
\jumpw{4}
\conf{\B\B\B\B\B\B\B\B\B\B\\ \B\B\B\O\B\X\B\B\B\B}

\item[(2)]

\quad
\conf{\B\B\B\O\B\X\B\B\B\B\\ \B\X\O\X\O\X\O\X\O\B\\ \X\O\X\O\X\O\X\O\X\O}
\jumpw{6}
\conf{\B\B\B\O\B\X\B\B\B\B\\ \B\B\B\X\O\X\O\B\B\B\\ \B\O\X\O\X\O\X\O\X\B}
\jumpw{4}
\conf{\B\B\B\O\B\B\B\B\B\B\\ \B\B\B\X\X\B\O\B\B\B\\ \B\B\O\O\X\B\X\O\X\B}
\jumpw{4}
\conf{\B\B\B\B\B\B\B\B\B\B\\ \B\B\B\B\B\B\B\B\B\B\\ \B\B\O\X\X\B\O\O\X\B}
\jumpw{4}
\conf{\B\B\B\B\B\B\B\B\B\B\\ \B\B\B\B\B\B\B\B\B\B\\ \B\B\B\B\O\B\X\B\B\B}

\item[(3)]

\quad
\conf{\B\B\B\B\O\B\X\B\B\B\\ \B\X\O\X\O\X\O\X\O\B\\ \X\O\X\O\X\O\X\O\X\O}
\jumpw{6}
\conf{\B\B\B\B\O\B\X\B\B\B\\ \B\B\B\X\O\X\O\B\B\B\\ \B\O\X\O\X\O\X\O\X\B}
\jumpw{4}
\conf{\B\B\B\B\B\B\X\B\B\B\\ \B\B\B\B\O\X\O\B\B\B\\ \B\B\O\O\X\O\X\X\B\B}
\jumpw{4}
\conf{\B\B\B\B\B\B\B\B\B\B\\ \B\B\B\B\B\B\O\B\B\B\\ \B\B\O\X\B\O\X\X\B\B}
\jumpw{4}
\conf{\B\B\B\B\B\B\B\B\B\B\\ \B\B\B\B\B\B\B\B\B\B\\ \B\B\B\O\B\X\B\B\B\B}
\end{itemize}

\subsection{Case OE: Odd Number of Rows, Even Number of Columns}
\label{ss_oe}

To extend Case EE from the previous section to handle an odd number of rows, we
could provide extra termination cases with three instead of
two rows for any step.
Because these steps are always final,
they may produce an arbitrary result configuration with one
or two stones.

However, as observed before, we only need to consider configurations
with three or five rows in Case OE
(any other configuration can be rotated into a Case EO).
It turns out that we can describe their reduction more easily
(and conform with all other cases) by first rotating them.
Thus,
the following reductions use the general approach from Section \ref{ss_general}
to reduce configurations with three or five columns
and an even number of rows.

\paragraph{Three Columns:}

\begin{itemize}
\item[(1)]

\quad
\conf{\X\O\X\\ \O\X\O}
\jumpw{2}
\conf{\B\B\X\\ \X\O\O}
\jumpw{2}
\conf{\B\B\B\\ \O\B\X}

\item[(2)]

\quad
\conf{\O\B\X\\ \X\O\X\\ \O\X\O}
\jumpw{2}
\conf{\O\B\B\\ \X\B\X\\ \O\X\O}
\jumpw{2}
\conf{\B\B\B\\ \O\B\X\\ \X\B\O}
\jumpw{2}
\conf{\B\B\B\\ \B\B\B\\ \O\B\X}
\end{itemize}

\paragraph{Five Columns:}

\begin{itemize}
\item[(1)]

\quad
\conf{\X\O\X\O\X\\ \O\X\O\X\O}
\jumpw{2}
\conf{\B\X\X\O\X\\ \B\O\O\X\O}
\jumpw{2}
\conf{\B\X\X\X\B\\ \B\O\O\O\B}
\jumpw{2}
\conf{\B\O\X\B\B\\ \B\B\O\X\B}
\jumpw{3}
\conf{\B\B\B\B\B\\ \B\B\O\B\B}

\item[(2)]

\quad
\conf{\B\B\O\B\B\\ \X\O\X\O\X\\ \O\X\O\X\O}
\jumpb{4}
\conf{\B\B\O\B\B\\ \B\X\X\X\B\\ \B\O\O\O\B}
\jumpb{3}
\conf{\B\B\B\B\B\\ \B\X\O\B\B\\ \B\O\X\B\B}
\jumpw{3}
\conf{\B\B\B\B\B\\ \B\B\B\B\B\\ \B\O\B\B\B}

\item[(3)]

\quad
\conf{\B\O\B\B\B\\ \X\O\X\O\X\\ \O\X\O\X\O}
\jumpb{4}
\conf{\B\O\B\B\B\\ \B\X\X\X\B\\ \B\O\O\O\B}
\jumpb{3}
\conf{\B\B\B\B\B\\ \B\O\B\X\B\\ \B\X\B\O\B}
\jumpw{2}
\conf{\B\B\B\B\B\\ \B\B\B\B\B\\ \B\O\B\X\B}

\item[(4)]

\quad
\conf{\B\O\B\X\B\\ \X\O\X\O\X\\ \O\X\O\X\O}
\jumpw{4}
\conf{\B\O\B\B\B\\ \B\X\X\X\O\\ \B\O\O\X\B}
\jumpw{2}
\conf{\B\B\B\B\B\\ \B\X\B\X\O\\ \B\O\O\X\B}
\jumpw{2}
\conf{\B\B\B\B\B\\ \B\B\B\O\B\\ \B\X\O\X\B}
\jumpw{3}
\conf{\B\B\B\B\B\\ \B\B\B\B\B\\ \B\B\O\B\B}
\end{itemize}

\subsection{Cases EO and OO: Odd Number of Columns}
\label{ss_o}

Finally we consider the case of an even or odd number of rows and 
an odd number of columns.
For each step, we give two variants, one reduction from two rows and one
reduction from three rows.
The latter case is applied only at the end of the reduction,
so it does not need to end with the same pattern of remainder stones.
Also, for an odd number of rows, the initial symmetrical removal of columns
from both ends of the rows in a step is done first for the final three-row
step, before any other reduction; this order is necessary because the
three-row symmetrical removal can start only with a White move.

The number of columns is at least seven.
Every step begins by reducing the two or three involved rows
down to a small number of columns.

\paragraph{Two Rows.}
First, we clobber a few stones to create the following configuration
in which the upper row has one more stone on the left side than
the lower row, and the lower row has one more stone on the right
side than the upper row:
\seq{
\conf{\X\O\\ \O\X}$\cdots$\conf{\O\X\\ \X\O}
\jumpw{3}
\conf{\B\O\\ \B\B}$\cdots$\conf{\O\X\\ \X\O}
\jumpb{3}
\conf{\B\O\\ \B\B}$\cdots$\conf{\B\B\\ \X\B}
}
Similar to Case EE,
we repeatedly apply the following reduction,
in each step removing six columns, three on each side:
\seq{
\conf{\O\X\O\X\\ \B\O\X\O}$\cdots$\conf{\X\O\X\B\\ \O\X\O\X}
\jumpw{3}
\conf{\O\X\O\X\\ \B\O\X\O}$\cdots$\conf{\X\B\B\B\\ \O\X\O\B}
\jumpb{3}
\conf{\B\X\O\X\\ \B\B\B\O}$\cdots$\conf{\X\B\B\B\\ \O\X\O\B}
\jumpw{2}
\conf{\B\B\X\X\\ \B\B\B\O}$\cdots$\conf{\X\B\B\B\\ \O\O\B\B}
\jumpw{4}
\conf{\B\B\B\X\\ \B\B\B\B}$\cdots$\conf{\B\B\B\B\\ \O\B\B\B}
}
We stop applying this reduction when the total number of columns is
just five, seven, or nine, so each row has 
four, six, or eight occupied columns,
depending on whether $m \equiv 1$, $0$, or $2 \pmod{3}$, respectively.

The resulting two-row configuration has either
(a) a black stone in the upper-left
and a white stone in the lower-right \new corner
$\Big(\conf{\X\O\X\\ \B\X\O}\cdots \conf{\X\O\B\\ \O\X\O}\Big)$,
or (b) vice versa
$\Big(\conf{\O\X\O\\ \B\O\X}\cdots \conf{\O\X\B\\ \X\O\X}\Big)$.
We will show reductions from both configurations.
It turns out that configuration (a) is more difficult
to handle because it is not always possible to end up with a
single stone (or pair of stones) on the bottom row.
In that case, we will make the last move parenthetical,
omitting it whenever this step is not the last.

Sometimes we also need to start from the configuration
(a$^\prime$) $\conf{\B\O\X\\ \O\X\O}\cdots \conf{\X\O\X\\ \O\X\B}$
or (b$^\prime$) $\conf{\B\X\O\\ \X\O\X}\cdots \conf{\O\X\O\\ \X\O\B}$
which are the mirror images of the configurations (a) and (b).
These starting points can be achieved by applying the reductions above
upside-down.

\paragraph{Three rows.}

First, we clobber a few stones to create the following configuration:
\seq{
\conf{\X\O\\ \O\X\\ \X\O}$\cdots$\conf{\O\X\\ \X\O\\ \O\X}
\jumpw{8}
\conf{\B\X\\ \B\B\\ \B\O}$\cdots$\conf{\O\B\\ \B\B\\ \X\B}
}
Then, we reduce long rows by four columns at a time
(not six as in the two-row reductions):
\seq{
\conf{\X\X\O\\ \B\O\X\\ \O\X\O}$\cdots$\conf{\O\X\O\\ \X\O\B\\ \O\X\X}
\jumpw{3}
\conf{\B\X\O\\ \B\B\X\\ \B\O\O}$\cdots$\conf{\O\X\O\\ \X\O\B\\ \O\X\X}
\jumpb{3}
\conf{\B\B\X\\ \B\B\B\\ \B\B\O}$\cdots$\conf{\O\X\O\\ \X\O\B\\ \O\X\X}
\jumpw{3}
\conf{\B\B\X\\ \B\B\B\\ \B\B\O}$\cdots$\conf{\O\O\B\\ \X\B\B\\ \O\X\B}
\jumpb{3}
\conf{\B\B\X\\ \B\B\B\\ \B\B\O}$\cdots$\conf{\O\B\B\\ \B\B\B\\ \X\B\B}
}
Note that we can also obtain the symmetric configuration
\conf{\B\B\O\\ \B\B\B\\ \B\B\X}$\cdots$\conf{\X\B\B\\ \B\B\B\\ \O\B\B}.
Because we cannot perform this reduction with Black starting, we must
perform this reduction at the very beginning of the entire algorithm,
before any other steps.

We stop this reduction when we have reached one of the three
configurations 
\conf{\X\X\O\\ \B\O\B\\ \O\X\X} or
\conf{\X\X\O\X\O\\ \B\O\X\O\B\\ \O\X\O\X\X} or
\conf{\X\X\O\X\O\X\O\\ \B\O\X\O\X\O\B\\ \O\X\O\X\O\X\X}.
\new The last configuration could be reduced further but
in some cases that would isolate the remaining stones from
the remainder stones of the rows above.

\paragraph{Reductions.}
Now we show how to reduce the configurations described above
following the general approach from Section \ref{ss_general}.
For the case of three rows, we only need to consider the following
two reductions in step (1):

\begin{itemize}
\item[(1)]

\quad
\conf{\X\X\O\\ \B\O\B\\ \O\X\X}
\jumpw{2}
\conf{\X\X\O\\ \B\O\B\\ \B\X\B}
\jumpw{3}
\conf{\B\X\B\\ \B\B\B\\ \B\O\B}

\item[(1$^\prime$)]

\quad
\conf{\X\X\O\X\O\\ \B\O\X\O\B\\ \O\X\O\X\X}
\jumpw{2}
\conf{\B\X\B\X\O\\ \B\O\X\O\B\\ \O\X\O\X\X}
\jumpw{2}
\conf{\B\X\B\O\B\\ \B\O\B\X\B\\ \O\X\O\X\X}
\jumpw{2}
\conf{\B\B\B\B\B\\ \B\X\B\O\B\\ \O\X\O\X\X}
\jumpw{2}
\conf{\B\B\B\B\B\\ \B\B\B\O\B\\ \B\X\O\X\X}
\jumpw{3}
\conf{\B\B\B\B\B\\ \B\B\B\B\B\\ \B\X\B\O\B}
\end{itemize}

\paragraph{Case 1: $m\equiv 1\pmod{3}$}

\ 

The initial configuration is of type (a)
for $m=13+12k$ columns
and of type (b) for $m=7+12k$ columns, for $k\ge 0$.

\begin{itemize}
\item[(1a)]

\quad
\conf{\X\O\X\O\B\\ \B\X\O\X\O}
\jumpw{2}
\conf{\X\O\X\O\B\\ \B\B\X\O\B}
\jumpw{2}
\conf{\B\X\O\B\B\\ \B\B\X\O\B}
\jumpw{2}
\conf{\B\B\X\B\B\\ \B\B\O\B\B}
\lparen
\move
\conf{\B\B\O\B\B\\ \B\B\B\B\B}
\rparen

\item[(1b)]

\quad
\conf{\O\X\O\X\B\\ \B\O\X\O\X}
\jumpw{2}
\conf{\B\O\X\B\B\\ \B\O\X\O\X}
\jumpw{2}
\conf{\B\B\O\B\B\\ \B\O\X\X\B}
\jumpw{3}
\conf{\B\B\B\B\B\\ \B\B\O\B\B}

\item[(2a)]

\quad
\conf{\B\B\X\B\B\\ \B\B\O\B\B\\ \X\O\X\O\B\\ \B\X\O\X\O}
\jumpw{2}
\conf{\B\B\X\B\B\\ \B\B\O\B\B\\ \B\X\X\O\B\\ \B\O\B\X\O}
\jumpw{2}
\conf{\B\B\B\B\B\\ \B\B\X\B\B\\ \B\O\X\O\B\\ \B\B\B\X\O}
\jumpw{2}
\conf{\B\B\B\B\B\\ \B\B\B\B\B\\ \B\B\X\O\B\\ \B\B\B\X\O}
\jumpw{2}
\conf{\B\B\B\B\B\\ \B\B\B\B\B\\ \B\B\B\X\B\\ \B\B\B\O\B}
\lparen
\move
\conf{\B\B\B\B\B\\ \B\B\B\B\B\\ \B\B\B\O\B\\ \B\B\B\B\B}
\rparen

\smallskip

\quad
\conf{\B\B\B\X\B\B\B\\ \B\B\B\O\B\B\B\\ \X\X\O\X\O\X\O\\ \B\O\X\O\X\O\B\\ \O\X\O\X\O\X\X}
\jumpw{4}
\conf{\B\B\B\X\B\B\B\\ \B\B\B\O\B\B\B\\ \B\B\B\X\O\X\O\\ \B\X\X\O\X\O\B\\ \B\O\O\X\O\X\X}
\jumpw{4}
\conf{\B\B\B\X\B\B\B\\ \B\B\B\O\B\B\B\\ \B\B\B\X\O\X\O\\ \B\B\X\O\X\O\B\\ \B\B\B\B\O\X\X}
\jumpw{4}
\conf{\B\B\B\X\B\B\B\\ \B\B\B\O\B\B\B\\ \B\B\B\X\O\O\B\\ \B\B\X\O\X\X\B\\ \B\B\B\B\B\B\B}
\jumpw{4}
\conf{\B\B\B\B\B\B\B\\ \B\B\B\X\B\B\B\\ \B\B\B\X\O\B\B\\ \B\B\B\X\O\B\B\\ \B\B\B\B\B\B\B}
\jumpw{4}
\conf{\B\B\B\B\B\B\B\\ \B\B\B\B\B\B\B\\ \B\B\B\B\B\B\B\\ \B\B\B\X\B\B\B\\ \B\B\B\B\B\B\B}

\item[(2b)]

\quad
\conf{\B\B\O\B\B\\ \O\X\O\X\B\\ \B\O\X\O\X}
\jumpb{3}
\conf{\B\B\B\B\B\\ \O\X\O\B\B\\ \B\O\X\X\B}
\jumpw{2}
\conf{\B\B\B\B\B\\ \O\X\B\B\B\\ \B\O\X\B\B}
\jumpw{3}
\conf{\B\B\B\B\B\\ \B\B\B\B\B\\ \B\O\B\B\B}

\smallskip

\quad
\conf{\B\B\O\B\B\\ \X\X\O\X\O\\ \B\O\X\O\B\\ \O\X\O\X\X}
\jumpb{4}
\conf{\B\B\O\B\B\\ \X\X\O\X\O\\ \B\X\B\O\B\\ \B\O\B\X\B}
\jumpb{4}
\conf{\B\B\O\B\B\\ \X\X\O\X\B\\ \B\O\B\B\B\\ \B\B\B\B\B}
\jumpw{5}
\conf{\B\B\B\B\B\\ \B\B\X\B\B\\ \B\B\B\B\B\\ \B\B\B\B\B}

\item[(3a)]

\quad
\conf{\B\B\B\X\B\\ \B\B\B\O\B\\ \X\O\X\O\B\\ \B\X\O\X\O}
\jumpw{2}
\conf{\B\B\B\X\B\\ \B\B\B\O\B\\ \B\X\O\O\B\\ \B\X\B\X\O}
\jumpw{2}
\conf{\B\B\B\B\B\\ \B\B\B\X\B\\ \B\O\B\O\B\\ \B\X\B\X\O}
\jumpw{2}
\conf{\B\B\B\B\B\\ \B\B\B\B\B\\ \B\B\B\X\B\\ \B\O\B\X\O}
\jumpw{2}
\conf{\B\B\B\B\B\\ \B\B\B\B\B\\ \B\B\B\B\B\\ \B\O\B\X\B}

\smallskip

\quad
\conf{\B\B\B\B\X\B\B\\ \B\B\B\B\O\B\B\\ \X\X\O\X\O\X\O\\ \B\O\X\O\X\O\B\\ \O\X\O\X\O\X\X}
\jumpw{4}
\conf{\B\B\B\B\X\B\B\\ \B\B\B\B\O\B\B\\ \B\B\B\X\O\X\O\\ \B\X\X\O\X\O\B\\ \B\O\O\X\O\X\X}
\jumpw{4}
\conf{\B\B\B\B\X\B\B\\ \B\B\B\B\O\B\B\\ \B\B\B\X\O\X\O\\ \B\B\X\O\X\O\B\\ \B\B\B\B\O\X\X}
\jumpw{4}
\conf{\B\B\B\B\X\B\B\\ \B\B\B\B\O\B\B\\ \B\B\B\X\O\O\B\\ \B\B\X\O\X\X\B\\ \B\B\B\B\B\B\B}
\jumpw{4}
\conf{\B\B\B\B\B\B\B\\ \B\B\B\B\X\B\B\\ \B\B\B\X\O\B\B\\ \B\B\B\X\O\B\B\\ \B\B\B\B\B\B\B}
\jumpw{4}
\conf{\B\B\B\B\B\B\B\\ \B\B\B\B\B\B\B\\ \B\B\B\X\B\B\B\\ \B\B\B\B\B\B\B\\ \B\B\B\B\B\B\B}

\item[(3b)]

\quad
\conf{\B\O\B\B\B\\ \O\X\O\X\B\\ \B\O\X\O\X}
\jumpb{3}
\conf{\B\B\B\B\B\\ \O\X\B\B\B\\ \B\O\X\O\X}
\jumpw{2}
\conf{\B\B\B\B\B\\ \B\O\B\B\B\\ \B\X\B\O\X}
\jumpw{2}
\conf{\B\B\B\B\B\\ \B\B\B\B\B\\ \B\O\B\X\B}

\smallskip

\quad
\conf{\B\O\B\B\B\\ \X\X\O\X\O\\ \B\O\X\O\B\\ \O\X\O\X\X}
\jumpb{4}
\conf{\B\O\B\B\B\\ \X\X\O\X\O\\ \B\X\B\O\B\\ \B\O\B\X\B}
\jumpb{4}
\conf{\B\O\B\B\B\\ \X\X\O\X\B\\ \B\O\B\B\B\\ \B\B\B\B\B}
\jumpb{5}
\conf{\B\B\B\B\B\\ \B\X\B\B\B\\ \B\B\B\B\B\\ \B\B\B\B\B}

\item[(4a)]

\quad
\conf{\B\O\B\X\B\\ \X\O\X\O\B\\ \B\X\O\X\O}
\jumpw{2}
\conf{\B\O\B\X\B\\ \B\X\X\O\B\\ \B\X\O\O\B}
\jumpw{2}
\conf{\B\B\B\B\B\\ \B\O\X\X\B\\ \B\X\O\O\B}
\jumpw{2}
\conf{\B\B\B\B\B\\ \B\B\O\X\B\\ \B\B\X\O\B}
\jumpw{2}
\conf{\B\B\B\B\B\\ \B\B\X\B\B\\ \B\B\O\B\B}
\lparen
\move
\conf{\B\B\B\B\B\\ \B\B\O\B\B\\ \B\B\B\B\B}
\rparen

\smallskip

\quad
\conf{\O\B\X\\ \X\X\O\\ \B\O\B\\ \O\X\X}
\jumpw{2}
\conf{\B\B\B\\ \O\X\X\\ \B\O\B\\ \O\X\X}
\jumpw{2}
\conf{\B\B\B\\ \B\X\B\\ \B\O\B\\ \O\X\X}
\jumpw{3}
\conf{\B\B\B\\ \B\X\B\\ \B\B\B\\ \B\O\B}

\item[(4b$^\prime$)]

\new In the two-rows case, 
we must reduce rows seven and eight starting with the mirrored standard
initial configuration.

\quad
\conf{\B\O\B\X\B\\ \B\X\O\X\O\\ \X\O\X\O\B}
\jumpw{2}
\conf{\B\O\B\B\B\\ \B\X\O\X\B\\ \X\O\X\O\B}
\jumpw{2}
\conf{\B\O\B\B\B\\ \B\X\B\X\B\\ \B\X\O\O\B}
\jumpw{2}
\conf{\B\B\B\B\B\\ \B\O\B\B\B\\ \B\X\O\X\B}
\jumpw{3}
\conf{\B\B\B\B\B\\ \B\B\B\B\B\\ \B\B\O\B\B}

\smallskip

\quad
\conf{\B\O\B\X\B\\ \X\X\O\X\O\\ \B\O\X\O\B\\ \O\X\O\X\X}
\jumpw{4}
\conf{\B\O\B\X\B\\ \X\X\O\X\O\\ \B\X\B\O\B\\ \B\O\B\X\B}
\jumpw{4}
\conf{\B\O\B\X\B\\ \X\X\O\X\B\\ \B\O\B\B\B\\ \B\B\B\B\B}
\jumpw{5}
\conf{\B\B\B\B\B\\ \B\O\B\X\B\\ \B\B\B\B\B\\ \B\B\B\B\B}
\end{itemize}

\paragraph{Case 2: $m\equiv 0\pmod{3}$}

\ 

The initial configuration is of type (a)
for $m=15+12k$ columns
and of type (b) for $m=9+12k$ columns, for $k\ge 0$.

\begin{itemize}
\item[(1a)]

\quad
\conf{\X\O\X\O\X\O\B\\ \B\X\O\X\O\X\O}
\jumpw{2}
\conf{\B\X\O\B\X\O\B\\ \B\X\O\X\O\X\O}
\jumpw{2}
\conf{\B\O\B\B\X\O\B\\ \B\X\X\B\O\X\O}
\jumpw{2}
\conf{\B\O\B\B\B\X\B\\ \B\X\X\B\O\O\B}
\jumpw{4}
\conf{\B\B\B\B\B\B\B\\ \B\B\O\B\X\B\B}

\item[(1b)]

\quad
\conf{\O\X\O\X\O\X\B\\ \B\O\X\O\X\O\X}
\jumpw{2}
\conf{\B\O\X\B\O\X\B\\ \B\O\X\O\X\O\X}
\jumpw{2}
\conf{\B\X\B\B\O\X\B\\ \B\O\O\B\X\O\X}
\jumpw{2}
\conf{\B\X\B\B\B\O\B\\ \B\O\O\B\X\X\B}
\jumpw{4}
\conf{\B\B\B\B\B\B\B\\ \B\B\X\B\O\B\B}

\item[(2a)]

\quad
\conf{\B\B\O\B\X\B\B\\ \X\O\X\O\X\O\B\\ \B\X\O\X\O\X\O}
\jumpw{2}
\conf{\B\B\O\B\X\B\B\\ \B\X\X\O\O\B\B\\ \B\X\O\X\O\X\O}
\jumpw{2}
\conf{\B\B\B\B\X\B\B\\ \B\B\O\B\O\B\B\\ \B\B\X\X\O\X\O}
\jumpw{2}
\conf{\B\B\B\B\B\B\B\\ \B\B\B\B\X\B\B\\ \B\B\O\X\O\X\O}
\jumpw{2}
\conf{\B\B\B\B\B\B\B\\ \B\B\B\B\X\B\B\\ \B\B\O\B\X\O\B}
\jumpw{2}
\conf{\B\B\B\B\B\B\B\\ \B\B\B\B\B\B\B\\ \B\B\O\B\X\B\B}

\smallskip

For three rows, this case is identical to Case 1(4b).

\item[(2b)]

\quad
\conf{\B\B\X\B\O\B\B\\ \O\X\O\X\O\X\B\\ \B\O\X\O\X\O\X}
\jumpw{2}
\conf{\B\B\B\B\O\B\B\\ \B\O\X\X\O\X\B\\ \B\O\X\O\X\O\X}
\jumpw{2}
\conf{\B\B\B\B\O\B\B\\ \B\O\X\B\X\X\B\\ \B\O\O\B\X\O\X}
\jumpw{2}
\conf{\B\B\B\B\B\B\B\\ \B\X\B\B\O\X\B\\ \B\O\O\B\X\O\X}
\jumpw{6}
\conf{\B\B\B\B\B\B\B\\ \B\B\B\B\B\B\B\\ \B\B\X\B\O\B\B}

\smallskip

For three rows, this case is symmetric to (4a) in Case 1
(with the mirrored initial configuration).
\end{itemize}

\paragraph{Case 3: $m\equiv 2\pmod{3}$}

\ 

The initial configuration is of type (a)
for $m=17+12k$ columns
and of type (b) for $m=11+12k$ columns, for $k\ge 0$.

\begin{itemize}
\item[(1a)]

\quad
\conf{\X\O\X\O\X\O\X\O\B\\ \B\X\O\X\O\X\O\X\O}
\jumpw{4}
\conf{\B\X\X\O\X\O\B\X\B\\ \B\O\B\X\O\X\O\O\B}
\jumpw{4}
\conf{\B\B\O\X\X\O\B\B\B\\ \B\B\B\B\O\X\O\X\B}
\jumpw{4}
\conf{\B\B\B\O\X\B\B\B\B\\ \B\B\B\B\O\X\B\B\B}
\jumpw{3}
\conf{\B\B\B\B\B\B\B\B\B\\ \B\B\B\B\O\B\B\B\B}

\item[(1b)]

\quad
\conf{\O\X\O\X\O\X\O\X\B\\ \B\O\X\O\X\O\X\O\X}
\jumpw{4}
\conf{\B\O\O\X\O\X\B\O\B\\ \B\X\B\O\X\O\X\X\B}
\jumpw{4}
\conf{\B\B\X\X\O\X\B\B\B\\ \B\B\B\O\X\O\O\B\B}
\jumpw{4}
\conf{\B\B\B\X\B\B\B\B\B\\ \B\B\B\O\X\O\B\B\B}
\jumpw{3}
\conf{\B\B\B\B\B\B\B\B\B\\ \B\B\B\O\B\B\B\B\B}

\item[(2a)]

\quad
\conf{\B\B\B\B\O\B\B\B\B\\ \X\O\X\O\X\O\X\O\B\\ \B\X\O\X\O\X\O\X\O}
\jumpb{4}
\conf{\B\B\B\B\B\B\B\B\B\\ \X\O\X\O\O\O\B\B\B\\ \B\X\O\X\O\X\O\X\B}
\jumpb{4}
\conf{\B\B\B\B\B\B\B\B\B\\ \X\O\X\O\O\O\B\B\B\\ \B\O\B\X\B\X\B\B\B}
\jumpb{4}
\conf{\B\B\B\B\B\B\B\B\B\\ \B\B\O\O\O\X\B\B\B\\ \B\B\B\X\B\B\B\B\B}
\lparen
\jumpb{4}
\conf{\B\B\B\B\B\B\B\B\B\\ \B\B\B\B\O\B\B\B\B\\ \B\B\B\B\B\B\B\B\B}
\rparen

\smallskip

\quad
\conf{\B\B\B\O\B\B\B\\ \X\X\O\X\O\X\O\\ \B\O\X\O\X\O\B\\ \O\X\O\X\O\X\X}
\jumpb{4}
\conf{\B\B\B\O\B\B\B\\ \B\X\B\X\O\X\O\\ \B\O\X\O\X\O\B\\ \O\X\X\B\B\O\X}
\jumpb{4}
\conf{\B\B\B\O\B\B\B\\ \B\B\B\X\O\X\O\\ \B\O\X\O\X\O\B\\ \B\B\X\B\B\X\B}
\jumpb{4}
\conf{\B\B\B\O\B\B\B\\ \B\B\B\X\O\O\B\\ \B\B\X\O\X\X\B\\ \B\B\B\B\B\B\B}
\jumpb{4}
\conf{\B\B\B\B\B\B\B\\ \B\B\B\B\O\B\B\\ \B\B\X\O\X\B\B\\ \B\B\B\B\B\B\B}
\jumpb{3}
\conf{\B\B\B\B\B\B\B\\ \B\B\B\B\B\B\B\\ \B\B\B\B\X\B\B\\ \B\B\B\B\B\B\B}

\item[(2b)]

\quad
\conf{\B\B\B\O\B\B\B\B\B\\ \O\X\O\X\O\X\O\X\B\\ \B\O\X\O\X\O\X\O\X}
\jumpb{6}
\conf{\B\B\B\O\B\B\B\B\B\\ \B\X\O\X\O\X\B\B\B\\ \B\B\B\O\X\O\X\O\B}
\jumpb{3}
\conf{\B\B\B\O\B\B\B\B\B\\ \B\B\X\X\O\B\B\B\B\\ \B\B\B\O\X\X\O\B\B}
\jumpw{4}
\conf{\B\B\B\B\B\B\B\B\B\\ \B\B\B\B\X\B\B\B\B\\ \B\B\B\B\O\X\O\B\B}
\jumpw{3}
\conf{\B\B\B\B\B\B\B\B\B\\ \B\B\B\B\B\B\B\B\B\\ \B\B\B\B\O\B\B\B\B}

\smallskip

For three rows, this case is identical to Case 1(3b).

\item[(3a)]

\quad
\conf{\B\B\O\O\O\X\B\B\B\\ \B\B\B\X\B\B\B\B\B\\ \X\O\X\O\X\O\X\O\B\\ \B\X\O\X\O\X\O\X\O}
\hspace{-8pt}
\jumpb{4}
\conf{\B\B\O\O\X\B\B\B\B\\ \B\B\B\X\B\B\B\B\B\\ \X\O\X\O\X\O\X\O\B\\ \B\B\O\B\B\X\O\X\O}
\hspace{-8pt}
\jumpb{4}
\conf{\B\B\O\O\X\B\B\B\B\\ \B\B\B\X\B\B\B\B\B\\ \B\B\X\O\X\O\O\O\B\\ \B\B\B\B\B\X\B\X\O}
\hspace{-8pt}
\jumpb{4}
\conf{\B\B\B\B\B\B\B\B\B\\ \B\B\B\O\B\B\B\B\B\\ \B\B\B\X\X\O\O\O\B\\ \B\B\B\B\B\X\B\X\O}
\hspace{-8pt}
\jumpb{6}
\conf{\B\B\B\B\B\B\B\B\B\\ \B\B\B\B\B\B\B\B\B\\ \B\B\B\B\B\O\B\B\B\\ \B\B\B\B\B\X\B\B\O}
\ensuremath{\bigg(}
\hspace{-11pt}
\move
\hspace{-5pt}
\conf{\B\B\B\B\B\B\B\B\B\\ \B\B\B\B\B\B\B\B\B\\ \B\B\B\B\B\X\B\B\B\\ \B\B\B\B\B\B\B\B\O}
\hspace{-5pt}
\ensuremath{\bigg)}

\smallskip

\quad
\conf{\O\O\O\X\B\\ \B\X\B\B\\ \B\X\X\O\\ \B\B\O\B\\ \B\O\X\X}
\jumpb{4}
\conf{\B\O\B\B\B\\ \B\X\B\B\\ \B\X\X\O\\ \B\B\O\B\\ \B\B\O\X}
\jumpb{4}
\conf{\B\B\B\B\B\\ \B\O\B\B\\ \B\X\O\B\\ \B\B\X\B\\ \B\B\B\B}
\jumpb{3}
\conf{\B\B\B\B\B\\ \B\B\B\B\\ \B\X\B\B\\ \B\B\B\B\\ \B\B\B\B}

\item[(3b)]

\new In the two-rows case,
we must reduce rows five and six starting with the mirrored standard
initial configuration
(we could also solve the standard configuration, but then we could
not continue with step (4b)).

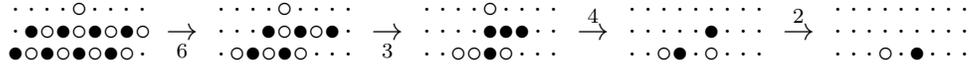
\quad
\conf{\B\B\B\B\O\B\B\B\B\\ \B\X\O\X\O\X\O\X\O\\ \X\O\X\O\X\O\X\O\B}
\jumpb{6}
\conf{\B\B\B\B\O\B\B\B\B\\ \B\B\B\X\O\X\O\X\B\\ \B\O\X\O\X\O\B\B\B}
\jumpb{3}
\conf{\B\B\B\B\O\B\B\B\B\\ \B\B\B\B\X\X\X\B\B\\ \B\B\O\O\X\O\B\B\B}
\jumpw{4}
\conf{\B\B\B\B\B\B\B\B\B\\ \B\B\B\B\B\X\B\B\B\\ \B\B\O\X\B\O\B\B\B}
\jumpw{2}
\conf{\B\B\B\B\B\B\B\B\B\\ \B\B\B\B\B\B\B\B\B\\ \B\B\B\O\B\X\B\B\B}

\smallskip

For three rows, this case is identical to Case 1(2b).

\item[(4a)]

\new In the two-rows case,
we must reduce rows seven and eight starting with the mirrored standard
initial configuration (because of the white single stone left over
at the right end of row six).

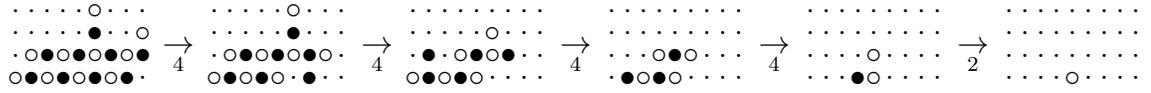
\quad
\conf{\B\B\B\B\B\O\B\B\B\\ \B\B\B\B\B\X\B\B\O\\ \B\O\X\O\X\O\X\O\X\\ \O\X\O\X\O\X\O\X\B}
\jumpb{4}
\conf{\B\B\B\B\B\O\B\B\B\\ \B\B\B\B\B\X\B\B\B\\ \B\O\X\O\X\O\X\O\B\\ \O\X\O\X\O\B\X\B\B}
\jumpb{4}
\conf{\B\B\B\B\B\B\B\B\B\\ \B\B\B\B\B\O\B\B\B\\ \B\X\B\O\X\O\X\B\B\\ \O\X\O\X\O\B\B\B\B}
\jumpb{4}
\conf{\B\B\B\B\B\B\B\B\B\\ \B\B\B\B\B\B\B\B\B\\ \B\B\B\O\X\O\B\B\B\\ \B\X\O\X\O\B\B\B\B}
\jumpb{4}
\conf{\B\B\B\B\B\B\B\B\B\\ \B\B\B\B\B\B\B\B\B\\ \B\B\B\B\O\B\B\B\B\\ \B\B\B\X\O\B\B\B\B}
\jumpb{2}
\conf{\B\B\B\B\B\B\B\B\B\\ \B\B\B\B\B\B\B\B\B\\ \B\B\B\B\B\B\B\B\B\\ \B\B\B\B\O\B\B\B\B}

\smallskip

In the three-rows case, we must reduce the number of columns
a little bit asymmetrically
(remove four additional columns on the left side) and then do the
following reduction.

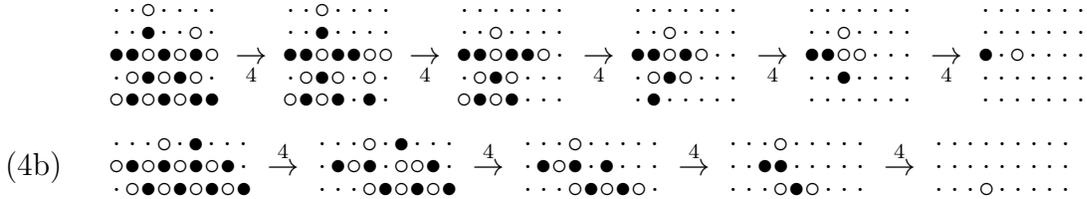
\quad
\conf{\B\B\O\B\B\B\B\\ \B\B\X\B\B\O\B\\ \X\X\O\X\O\X\O\\ \B\O\X\O\X\O\B\\ \O\X\O\X\O\X\X}
\jumpb{4}
\conf{\B\B\O\B\B\B\B\\ \B\B\X\B\B\B\B\\ \X\X\O\X\X\O\O\\ \B\O\X\O\B\O\B\\ \O\X\O\X\B\X\B}
\jumpb{4}
\conf{\B\B\B\B\B\B\B\\ \B\B\O\B\B\B\B\\ \X\X\O\X\X\O\B\\ \B\O\X\O\B\B\B\\ \O\X\O\X\B\B\B}
\jumpb{4}
\conf{\B\B\B\B\B\B\B\\ \B\B\O\B\B\B\B\\ \X\X\O\X\O\B\B\\ \B\O\X\O\B\B\B\\ \B\X\B\B\B\B\B}
\jumpb{4}
\conf{\B\B\B\B\B\B\B\\ \B\B\O\B\B\B\B\\ \X\X\O\O\B\B\B\\ \B\B\X\B\B\B\B\\ \B\B\B\B\B\B\B}
\jumpb{4}
\conf{\B\B\B\B\B\B\B\\ \B\B\B\B\B\B\B\\ \X\B\O\B\B\B\B\\ \B\B\B\B\B\B\B\\ \B\B\B\B\B\B\B}

\item[(4b)]

\quad
\conf{\B\B\B\O\B\X\B\B\B\\ \O\X\O\X\O\X\O\X\B\\ \B\O\X\O\X\O\X\O\X}
\jumpw{4}
\conf{\B\B\B\O\B\X\B\B\B\\ \B\X\O\X\B\O\O\X\B\\ \B\B\B\O\X\O\X\O\X}
\jumpw{4}
\conf{\B\B\B\O\B\B\B\B\B\\ \B\X\O\X\B\X\B\B\B\\ \B\B\B\O\X\O\X\O\B}
\jumpw{4}
\conf{\B\B\B\O\B\B\B\B\B\\ \B\B\X\X\B\B\B\B\B\\ \B\B\B\O\X\O\B\B\B}
\jumpw{4}
\conf{\B\B\B\B\B\B\B\B\B\\ \B\B\B\B\B\B\B\B\B\\ \B\B\B\O\B\B\B\B\B}

\smallskip

For three rows, this case is identical to Case 1(4b).
\end{itemize}

\section{NP-Completeness of 1-Reducibility}
\label{s_npc}

In this section we consider arbitrary initial Clobber positions
that do not need to have a rectangular shape or the
alternating checkerboard placement of the stones.
We show that then the following problem is NP-complete.

\bigskip

\fbox{
\begin{minipage}{.8\textwidth}
{\bf Problem \Clobber:}

Given an arbitrary initial Clobber configuration,
decide whether we can reduce it to a single stone.
\end{minipage}
}

\bigskip

The proof is by reduction from the Hamiltonian circuit problem
in grid graphs.
A {\em grid graph} is a finite graph embedded in the Euclidean plane
such that the vertices have integer coordinates
and two vertices are connected by an edge if and only if their
Euclidean distance is equal to one.

\bigskip

\fbox{
\begin{minipage}{.8\textwidth}
{\bf Problem \Hamgrid:}

Decide whether a given grid graph has a Hamiltonian circuit.
\end{minipage}
}

\bigskip
\bigskip

Itai \etal\ proved that \Hamgrid\ is NP-complete
\cite[Theorem 2.1]{ItPaSz82}.

\begin{theorem}
\label{thm_clobber_npc}
\Clobber\ is NP-complete.
\end{theorem}

\begin{proof}
We first observe that \Clobber\ is indeed in NP, because
we can easily check in polynomial time whether a proposed
solution (which must have only $n-1$ moves)
reduces the given initial configuration to a single stone.

We prove the NP-completeness by reduction from \Hamgrid.
Let $G$ be an arbitrary grid graph with $n$ nodes,
embedded in the Euclidean plane.
Let $v$ be a node of $G$ with maximum $y$-coordinate,
and among all such nodes the node with maximum $x$-coordinate.
If $v$ does not have a neighbor to the left then $G$ cannot have
a Hamiltonian circuit.
So assume there is a left neighbor $w$ of $v$.
Note that $v$ has degree two and therefore
any Hamiltonian circuit in $G$ must use the edge
$(v,w)$.

  \begin{figure}[htbp]
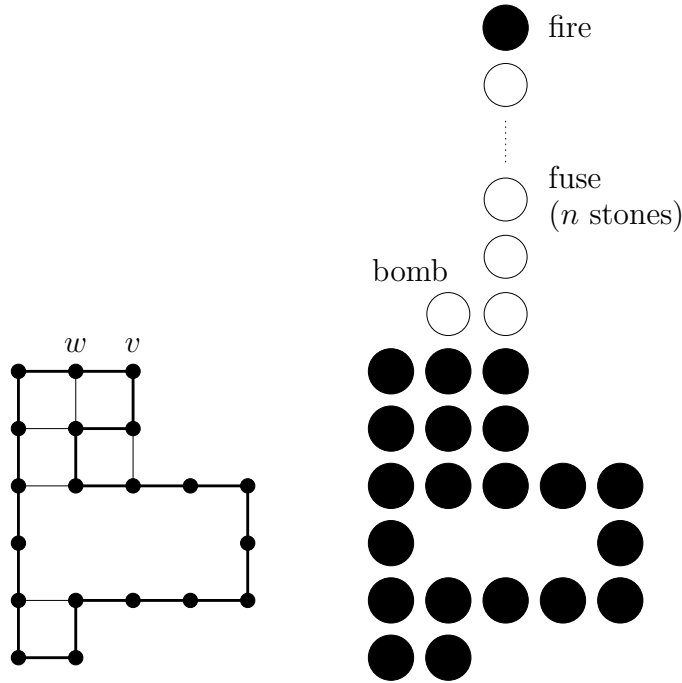

  \begin{center}
  \input npc.tex
  \caption{\label{fig_npc} An $n$-node grid graph with Hamiltonian circuit and the
corresponding Clobber configuration that can be reduced to
a single stone.}
  \end{center}
  \end{figure}%

Then we construct the following Clobber configuration
(see Figure~\ref{fig_npc}).
We put a black stone on each node of $G$.
We place a single white stone just above $w$, the {\em bomb}.
We place a vertical chain of $n$ white stones above $v$,
the {\em fuse}, and another
single black stone, the {\em fire}, on top of the fuse.
Altogether we have placed $n+1$ white and $n+1$ black stones,
so this is a legal Clobber configuration.

If $G$ has a Hamiltonian circuit $C$ then the bomb can clobber all
black nodes of $G$, following $C$ starting in $w$ and ending in
$v$ after $n$ rounds.
At the same time, the black fire can clobber the $n$ stones of the fuse
and end up just above $v$ after $n$ rounds.
But then in a last step the bomb can clobber the fire,
leaving a single stone on the board. 

On the other hand, if the initial configuration can be reduced to
a single stone then White cannot move any stone on the fuse
(because that would disconnect the black fire from the stones
on $G$), so it must move the bomb until Black has clobbered the
fuse.
But that takes $n$ steps, so White must in the meanwhile
clobber all $n$ black stones of $G$, that is,
it must walk along a Hamiltonian circuit in $G$.
\end{proof}

\section{Conclusions}
\label{s_conclusions}

We have seen that reducing \new a given Clobber configuration
to the minimum number of stones is polynomially
solvable for \new rectangular checkerboard configurations,
and is NP-hard for general configurations.
What about \new non-rectangular checkerboard configurations
and rectangular non-checkerboard configurations?

We have also seen a lower bound on the number of stones
to which a configuration can be reduced that is based on
the number of stones plus the number of stones on squares
of different color.
It would be interesting to identify other structural
parameters of a configuration that influence reducibility.

\bibliographystyle{plain}
\bibliography{clobber}

\begin{thebibliography}{1}

\bibitem{dag02}
E.~D. Demaine, R.~Fleischer, A.~Fraenkel, and R.~Nowakowski (organizers).
\newblock Dagstuhl {S}eminar on {A}lgorithmic {C}ombinatorial {G}ame {T}heory.
\newblock Seminar no.~02081, report no.~334, March 17--22, 2002.

\bibitem{Gro02}
J.~P. Grossmann, March 2002.
\newblock Private communication.

\bibitem{ItPaSz82}
A.~Itai, C.~H. Papadimitriou, and J.~L. Szwarcfiter.
\newblock Hamilton paths in grid graphs.
\newblock {\em SIAM Journal on Computing}, 11(4):676--686, 1982.

\bibitem{clobber_www}
D.~Wolfe.
\newblock Clobber {R}esearch, 2002.
\newblock \url{http://www.gac.edu/~wolfe/games/clobber}.

\end{thebibliography}

\appendix
\section{Small Cases}
\label{appendix}

Our proof of Theorem \ref{thm_2dclob_up} requires us to verify reducibility for
all instances with $2 \leq n, m \leq 6$.
This fact can be checked easily by a
computer, but for completeness we give the reductions here.
By symmetry, we only need to show the cases with $n\le m$.
Reductions of $2\times 3$, $2\times 5$,
$3\times 4$, $3\times 6$,
$4\times 5$, and $5\times 6$ boards are already given in
Section~\ref{ss_oe}.
Eight more small boards remain.
We assume White moves first.

\begin{description}
\item[$\bf 2{\times}2$:]

\quad
\conf{\X\O\\ \O\X}
\move
\conf{\X\B\\ \O\O}
\move
\conf{\B\B\\ \X\O}
\move
\conf{\B\B\\ \O\B}

\item[$\bf 2{\times}4$:]

\quad
\conf{\X\O\X\O\\ \O\X\O\X}
\move
\conf{\X\O\X\B\\ \O\X\O\O}
\move
\conf{\X\O\B\B\\ \O\X\X\O}
\move
\conf{\X\O\B\B\\ \O\X\O\B}
\move
\conf{\B\O\B\B\\ \X\X\O\B}
\move
\conf{\B\B\B\B\\ \X\O\O\B}
\move
\conf{\B\B\B\B\\ \B\X\O\B}
\move
\conf{\B\B\B\B\\ \B\O\B\B}

\item[$\bf 2{\times}6$:]

\quad
\conf{\X\O\X\O\X\O\\ \O\X\O\X\O\X}
\jumpw{3}
\conf{\X\O\X\O\B\B\\ \O\X\O\X\O\B}
\jumpb{3}
\conf{\B\B\X\O\B\B\\ \B\X\O\X\O\B}
\jumpw{2}
\conf{\B\B\B\B\B\B\\ \B\X\X\O\O\B}
\jumpw{2}
\conf{\B\B\B\B\B\B\\ \B\B\X\B\O\B}

\item[$\bf 3{\times}3$:]

\quad
\conf{\X\O\X\\ \O\X\O\\ \X\O\X}
\jumpw{2}
\conf{\O\B\X\\ \X\B\O\\ \X\O\X}
\jumpw{3}
\conf{\B\B\B\\ \B\B\B\\ \O\B\X}

\item[$\bf 3{\times}5$:]

\quad
\conf{\X\O\X\O\X\\ \O\X\O\X\O\\ \X\O\X\O\X}
\jumpw{2}
\conf{\X\O\X\O\X\\ \B\B\O\X\O\\ \O\X\X\O\X}
\jumpw{2}
\conf{\B\X\X\O\X\\ \B\B\O\X\O\\ \B\O\X\O\X}
\jumpw{2}
\conf{\B\B\X\O\X\\ \B\B\B\X\O\\ \B\O\X\O\X}
\jumpw{2}
\conf{\B\B\B\X\O\\ \B\B\B\X\B\\ \B\O\X\O\X}
\jumpw{2}
\conf{\B\B\B\O\B\\ \B\B\B\X\B\\ \B\O\X\X\B}
\jumpw{3}
\conf{\B\B\B\B\B\\ \B\B\B\B\B\\ \B\X\B\O\B}

\item[$\bf 4{\times}4$:]

We reduce the upper two rows as in the $2\times 4$ board, then
Black moves next:

\quad
\conf{\B\O\B\B\\ \X\O\X\O\\ \O\X\O\X}
\move
\conf{\B\O\B\B\\ \B\X\X\O\\ \O\X\O\X}
\move
\conf{\B\B\B\B\\ \B\O\X\O\\ \O\X\O\X}
\move
\conf{\B\B\B\B\\ \B\X\B\O\\ \O\X\O\X}
\move
\conf{\B\B\B\B\\ \B\X\B\O\\ \B\O\O\X}
\move
\conf{\B\B\B\B\\ \B\B\B\O\\ \B\X\O\X}
\move
\conf{\B\B\B\B\\ \B\B\B\B\\ \B\X\O\O}
\move
\conf{\B\B\B\B\\ \B\B\B\B\\ \B\B\X\O}
\move
\conf{\B\B\B\B\\ \B\B\B\B\\ \B\B\O\B}

\item[$\bf 4{\times}6$:]

We reduce the upper two rows as in the $2\times 6$ board, then
White moves next:

\conf{\B\B\X\B\O\B\\ \X\O\X\O\X\O\\ \O\X\O\X\O\X}
\jumpw{2}
\conf{\B\B\B\B\O\B\\ \X\B\X\O\X\O\\ \O\X\O\X\O\X}
\jumpw{2}
\conf{\B\B\B\B\B\B\\ \X\B\X\O\O\X\\ \O\X\O\X\O\B}
\jumpw{2}
\conf{\B\B\B\B\B\B\\ \X\B\X\O\X\B\\ \O\O\B\X\O\B}
\jumpw{2}
\conf{\B\B\B\B\B\B\\ \X\B\B\X\O\B\\ \O\O\B\X\B\B}
\jumpw{4}
\conf{\B\B\B\B\B\B\\ \B\B\B\B\B\B\\ \B\X\B\O\B\B}

\item[$\bf 6{\times}6$:]

We reduce the upper four rows as in the $4\times 6$ board, then
White moves next:

\conf{\B\X\B\O\B\B\\ \X\O\X\O\X\O\\ \O\X\O\X\O\X}
\jumpw{2}
\conf{\B\B\B\O\B\B\\ \O\X\X\O\X\O\\ \B\X\O\X\O\X}
\jumpw{2}
\conf{\B\B\B\O\B\B\\ \B\O\X\X\X\O\\ \B\X\O\B\O\X}
\jumpw{2}
\conf{\B\B\B\B\B\B\\ \B\X\X\O\X\O\\ \B\B\O\B\O\X}
\jumpw{2}
\conf{\B\B\B\B\B\B\\ \B\B\X\B\X\O\\ \B\B\O\B\O\X}
\jumpw{2}
\conf{\B\B\B\B\B\B\\ \B\B\X\B\B\B\\ \B\B\O\B\X\O}
\jumpw{2}
\conf{\B\B\B\B\B\B\\ \B\B\B\B\B\B\\ \B\B\X\B\O\B}
\end{description}

\end{document}